\documentclass[fleqn,10pt]{wlscirep}
\usepackage{makecell}

\title{Micro-mobility dispatch optimization via quantum annealing incorporating historical data}

\author[1,2,*]{Takeru Goto}
\author[1,3,4]{Masayuki Ohzeki}

\affil[1]{Graduate School of Information Sciences, Tohoku University, Miyagi, 980-8579, Japan}
\affil[2]{Innovative Research Excellence, Honda R\&D Co., Ltd., Tokyo, 107-6238, Japan}
\affil[3]{Department of Physics, Institute of Science Tokyo, Tokyo, 152-8551, Japan}
\affil[4]{Sigma-i Co., Ltd., 108-0075, Tokyo, Japan}
\affil[*]{goto.takeru.s8@dc.tohoku.ac.jp}


\begin{abstract}
This paper proposes a novel dispatch formulation for micro-mobility vehicles using a Quantum Annealer (QA). In recent years, QA has gained increasing attention as a high-performance solver for combinatorial optimization problems. Meanwhile, micro-mobility services have been rapidly developed as a promising means of realizing efficient and sustainable urban transportation. In this study, the dispatch problem for such micro-mobility services is formulated as a Quadratic Unconstrained Binary Optimization (QUBO) problem, enabling efficient solving through QA.
Furthermore, the proposed formulation incorporates historical usage data to enhance operational efficiency. Specifically, customer arrival frequencies and destination distributions are modeled into the QUBO formulation through a Bayesian approach, which guides the allocation of vacant vehicles to designated stations for waiting and charging.
Simulation experiments are conducted to evaluate the effectiveness of the proposed method, with comparisons to conventional formulations such as the vehicle routing problem. Additionally, the performance of QA is compared with that of classical solvers to reveal its potential advantages for the proposed dispatch formulation. The effect of reverse annealing on improving solution quality is also investigated.
\end{abstract}
\begin{document}

\flushbottom
\maketitle

\thispagestyle{empty}

\section*{Introduction}
In recent years, taxi dispatch systems and autonomous vehicle services have rapidly evolved around the world. In parallel, micro-mobility solutions, such as shared electric scooters and small autonomous vehicles, have gained increasing attention as a means to enhance transportation efficiency and environmental sustainability in densely populated urban areas. These systems offer flexible, short-distance mobility options while reducing traffic congestion and carbon emissions. With ongoing advances in automation and connectivity\cite{honda}, future micro-mobility fleets are envisioned to operate autonomously without human intervention. Accordingly, efficient and adaptive dispatch algorithms play a critical role in ensuring both time efficiency and energy optimization in such systems.

Traditionally, dispatch and routing problems in transportation and logistics have been formulated as variants of Vehicle Routing Problems (VRPs)\cite{dantzig1959truck}. The VRP is an NP-hard class\cite{lenstra1981complexity} of combinatorial optimization problems and has been widely studied due to its broad applicability\cite{laporte2009fifty}. Numerous extensions of the VRP have been proposed to accommodate real-world constraints. For instance, the Capacitated VRP (CVRP) imposes vehicle load limits\cite{ralphs2003capacitated}, making it suitable for ride-sharing and delivery services; the VRP with Time Windows\cite{solomon1987algorithms,liu2023systematic} (VRPTW) enforces customer-specific pickup or delivery times; and the Distance-Constrained VRP\cite{laporte1984two,kek2008distance} (DCVRP) incorporates constraints related to fuel and battery levels\cite{kucukoglu2021electric}. These problems are typically formulated as mathematical optimization models, and a wide range of solution methods have been developed to solve them efficiently.

On the other hand, general-purpose metaheuristic approaches such as genetic algorithms and simulated annealing have demonstrated strong versatility for various combinatorial problems. More recently, quantum computing paradigms have emerged as potential alternatives for solving large-scale NP-hard problems\cite{kadowaki1998quantum,farhi2014quantum,minami2025generative}. Among them, Quantum Annealer\cite{kadowaki1998quantum} (QA) has attracted substantial attention as it can directly handle Quadratic Unconstrained Binary Optimization (QUBO) formulations, which can represent a wide variety of combinatorial problems\cite{lucas2014ising}. QA gradually transforms a quantum system from an initial superposition state to the ground state of a problem-specific Hamiltonian by adiabatically reducing the strength of the transverse field\cite{morita2008mathematical, ohzeki2011quantum}. Commercial quantum annealers developed by D-Wave Systems\cite{mcgeochDWaveAdvantageSystem} have enabled practical experimentation with QA in diverse domains such as traffic optimization\cite{neukart2017traffic,inoue2021traffic,de2023formulation,haba2025routing}, logistics\cite{weinberg2023supply}, finance\cite{rosenberg2016solving, orus2019forecasting}, manufacturing\cite{ohzeki2019control,haba2022travel,quang2025quantum}, materials science\cite{Yonaga2022, tanaka2023virtual}, marketing\cite{nishimura2019item}, and machine learning\cite{neven2012qboost, khoshaman2018quantum, goto2025online}. To overcome the limitations of current hardware, such as restricted qubit connectivity and scalability, several relaxation and quantum-classical hybrid approaches have been proposed\cite{ohzeki2020breaking,yonaga2020solving,takabayashi2025subgradient}. Additionally, Reverse Annealing (RA), which refines solutions from high-quality initial states by scheduling the transverse field, has been studied theoretically\cite{Yamashiro2019} and applied to practical optimization problems\cite{venturelli2019reverse, haba2022travel}.

In the context of transportation optimization, several studies\cite{irie2019quantum, borowski2020new, feld2019hybrid,cattelan2024modeling} have applied QA to VRP formulation and related problems. For example, comprehensive QUBO formulations incorporating constraints such as time, vehicle state and capacity have been proposed\cite{irie2019quantum}. Other works\cite{borowski2020new,feld2019hybrid} introduced decomposition or partitioning schemes to reduce problem size and enhance scalability. More recently, QA-based approaches have been extended to ride-hailing and ride-pooling settings, where multiple passengers with intermediate pickups and drop-offs are handled simultaneously\cite{cattelan2024modeling}.

However, the suitability of VRP-based formulations for micro-mobility dispatch remains unclear. In contrast to conventional fleet operations, micro-mobility systems are characterized by highly dynamic, stochastic demand and frequent vehicle reassignments during operation. When customer requests arrive continuously, long-horizon route planning becomes less relevant, as vehicles must be reassigned adaptively rather than following predetermined sequences. Moreover, since micro-mobility vehicles typically serve a single passenger, the classical VRP capacity constraints play only a limited role. Instead, efficient management of idle vehicle allocation, for instance, determining how many vehicles should wait at each charging station, is crucial to ensure rapid response to future demand.

To address these challenges, we propose a novel formulation for the Micro-Mobility Dispatch Problem (hereafter referred to as MMDP). The proposed model estimates the optimal distribution of idle vehicles across stations based on historical usage data through Bayesian modeling, prioritizes customers with longer waiting times, and minimizes total travel time. By intentionally omitting service sequence variables, the formulation significantly reduces the number of binary variables, thereby improving scalability for real-time optimization. We first evaluate the validity of the proposed formulation by comparing it with two classical baselines: a distance-based greedy heuristic and a VRP-based QUBO formulation modified for MMDP. To rigorously validate the fundamental efficiency of our new formulation, we construct a grid-based simulation environment. Subsequently, we implement the MMDP formulation on the D-Wave Advantage to examine the characteristics of QA in this context. We compare its performance with that of the Gurobi Optimizer and further investigate the effect of RA on our scheme.

\section*{Methods}
\subsection*{Problem setting}

\begin{figure}[t]
\centering
\includegraphics[width=17cm]{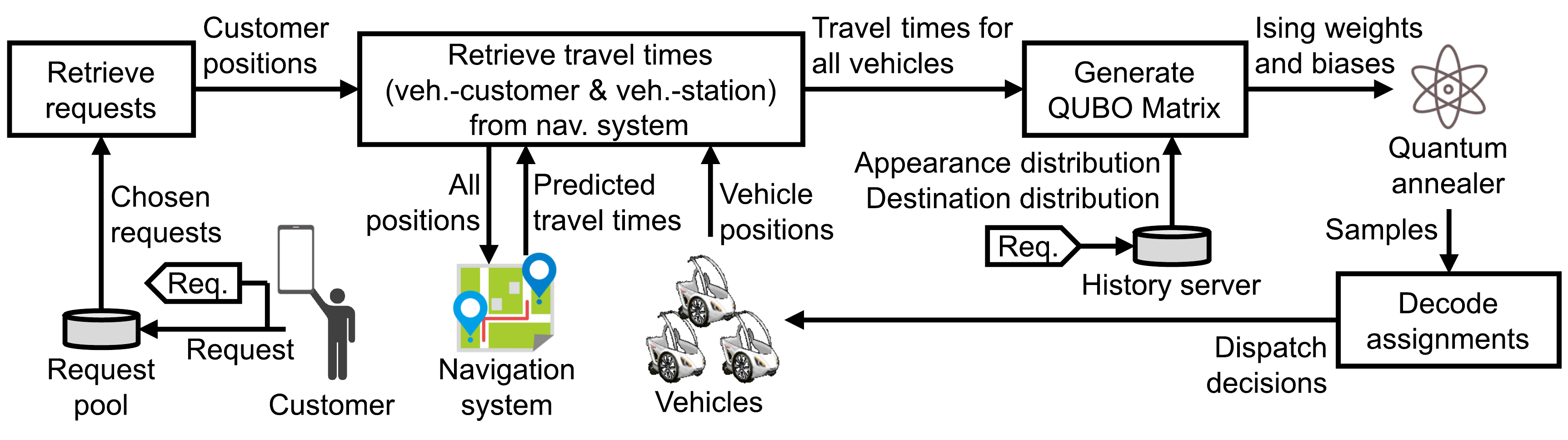}
\caption{Overview of the proposed micro-mobility dispatch system. Customer demand triggers the dispatch process, in which the system predicts travel times for vehicle–customer and vehicle–station pairs. The QUBO matrix is then constructed using the predicted travel times together with historical distribution data, and the resulting QUBO is solved by a quantum annealer.}
\label{fig:system}
\end{figure}

We consider a micro-mobility dispatch system composed of autonomous single-passenger vehicles operating within an urban area. The service area comprises multiple designated stations that can accommodate a sufficient number of vehicles for charging and standby. Customers request rides through a web-based or mobile application, after which the dispatch system determines whether each available vehicle should be assigned to a customer or repositioned to a specific station. All vehicles are continuously assigned either to a customer or to a station. If a vehicle is occupied, it proceeds to its next destination immediately after completing the current trip.

Figure $\ref{fig:system}$ presents an overview of the proposed dispatch system, including the data flow from customer demand to dispatch decisions. The system is assumed to have access to historical operational data, such as the spatiotemporal distribution of customer demand, destination patterns, and estimated travel times provided by an external navigation service. Using both historical and real-time information, the dispatch controller can operate adaptively and efficiently in response to real-time customer requests.

\subsection*{QUBO formulation}
The total number of vehicles is denoted by $N$. The current position and destination of vehicle $v_i$ are denoted by $x_{v_i}$ and $x_{d_i}$, respectively. If $x_{d_i} \neq x_{v_i}$, the vehicle is occupied and transporting a customer; otherwise, the vehicle is vacant.
Let $n_s$ and $n_c$ denote the number of stations and customers, respectively, both considered as dispatch targets. Since our approach assigns at most one customer to each vehicle in a single planning step, at most $N$ customers can be handled. Therefore, when the number of customers exceeds $N$, we extract the $N$ earliest-received customers and set $n_c=N$ in the subsequent formulation. The positions of station $s_j$ and customer $c_k$ are denoted by $x_{s_j}$ and $x_{c_k}$, respectively.
Two types of binary variables are introduced: $\sigma_{v,s} \in \{0,1\}^{Nn_s}$ and $\sigma_{v,c} \in \{0,1\}^{Nn_c}$. If vehicle $v_i$ is dispatched to station $s_j$, the corresponding variable $\sigma_{v_i,s_j}$ in $\sigma_{v,s}$ is set to one. Similarly, $\sigma_{v_i,c_k} = 1$ indicates that customer $c_k$ is assigned to vehicle $v_i$.
Although occupied vehicles may receive new assignments, they travel to the newly assigned destination only after reaching the destination of their current passenger.

Each vehicle can be assigned to only one target. This constraint is expressed by the Hamiltonian
\begin{equation}
H_{A_0}=\sum_{i}\left(1-\sum_j{\sigma_{v_i,s_j}-\sum_k{\sigma_{v_i,c_k}}}\right)^2.
\end{equation}
Each customer must be assigned to exactly one vehicle, formulated as
\begin{equation}
H_{A_1}=\sum_{k}\left(1-\sum_{i}\sigma_{v_i,c_k}\right)^2.
\end{equation}
The total travel time cost is represented by the Hamiltonian
\begin{equation}
H_{B_0}=\frac{1}{t_{\mathrm{avg}}}\left\{\sum_{i,j}t(x_{v_i},x_{d_i},x_{s_j})\sigma_{v_i,s_j}+\sum_{i,k}t(x_{v_i},x_{d_i},x_{c_k})\sigma_{v_i,c_k}\right\},\label{eq:hb0}
\end{equation}
where
\begin{equation}
t_{\mathrm{avg}} = \frac{\sum_{i,j}t(x_{v_i},x_{d_i},x_{s_j})+\sum_{i,k}t(x_{v_i},x_{d_i},x_{c_k})}{N(n_s+n_c)}.
\end{equation}
Here, $t(x_0, x_1, x_2)$ denotes the travel time from $x_0$ to $x_2$ via $x_1$. The division by $t_{\mathrm{avg}}$ normalizes the scale of $H_{B_0}$ relative to the other Hamiltonians.
To encourage more vehicles to be allocated to stations near areas with higher customer appearance frequencies, we use the Hamiltonian
\begin{equation}
H_{B_1}=\sum_j\left(\tau_j-\sum_i\sigma_{v_i,s_j}\right)^2,
\end{equation}
where $\tau_j$ represents the desirable number of vehicles allocated to station $s_j$.
Finally, the total Hamiltonian is defined as
\begin{equation}
H=H_{A_0}+H_{A_1}+B_0H_{B_0}+B_1H_{B_1},\label{eq:total}
\end{equation}
where $B_i$ are weights for the cost sub-Hamiltonians.
In the following subsection, we define and derive $\tau_j$ using historical usage data and the predicted travel times to destinations.

\subsection*{Desirable number of vehicles for each station}
We define $\tau_j$ as the expected number of new customer requests that will arise during the expected travel time to the station ($t_{s_j}$):
\begin{equation}
\tau_j \triangleq t_{s_j}f_cP(s_j),
\end{equation} 
where $f_c$ is the frequency of customer appearances, and $P(s_j)$ is the selection probability of vehicles allocated to station $s_j$.
Since there is a time gap between the dispatch decision and the vehicle arrival, the system should preemptively allocate vehicles to cover the demand generated during this interval. By setting the target inventory $\tau_j$ equal to this expected demand, we aim to maintain a supply-demand equilibrium.
To enhance operational efficiency, vacant vehicles should be distributed to stations located near areas with high customer appearance frequencies. Therefore, we formulate $P(s_j)$ by marginalizing the conditional selection probability over the customer location $x_{c_k}$:
\begin{equation}
P(s_j)=\int_{x_{c_k}}p(x_{c_k})P(s_j \mid x_{c_k})dx_{c_k}.\label{eq:Psj}
\end{equation}
Inside the integral, assuming a uniform prior (i.e. all stations are intrinsically equally selectable before considering distance), the conditional probability $P(s_j \mid x_{c_k})$ is directly proportional to the likelihood as $P(s_j \mid x_{c_k}) \propto p(x_{c_k} \mid s_j)$.
Furthermore, $p(x_{c_k} \mid s_j)$ is approximated by an exponential distribution that decreases with the travel time $t(x_{s_j},x_{c_k})$ as:
\begin{equation}
p(x_{c_k} \mid s_j) \approx \frac{1}{\theta_c}e^{-t(x_{s_j},x_{c_k})/{\theta_c}},\label{eq:pxc}
\end{equation}
where $\theta_c$ is the average dispatch time from stations to customers and $t(x_0,x_1)$ denotes the direct travel time from $x_0$ to $x_1$.
This approximation models the distance decay effect, a standard concept in spatial interaction models. Since the dispatch selection depends on temporal proximity, the probability is modeled to decay with travel time rather than physical distance.
Substituting these assumptions and normalizing the probability into Eq. (\ref{eq:Psj}), we obtain:
\begin{equation}
P(s_j)=\int_{x_{c_k}}p(x_{c_k})\frac{e^{-t(x_{s_j},x_{c_k})/{\theta_c}}}{\sum_{s'_j}e^{-t(x_{s'_j},x_{c_k})/{\theta_c}}}dx_{c_k}.
\end{equation}
Both $p(x_{c_k})$ and $f_c$ are obtained from the historical data.
The expected travel time $t_{s_j}$ for station $s_j$ is formulated as:
\begin{equation}
t_{s_j} = \sum_{\sigma_{v,s_j} \neq \boldsymbol{0}}\frac{P(\sigma_{v,s_j} \mid x_c,x_v,x_d)}{Z_{\sigma_{v,s_j} \neq \boldsymbol{0}}} 
\frac{\sum_{\sigma_{v_i,s_j} \in \sigma_{v,s_j}}\sigma_{v_i,s_j}t(x_{v_i},x_{d_i},x_{s_j})}{\sum_{\sigma_{v_i,s_j} \in \sigma_{v,s_j}}\sigma_{v_i,s_j}},\label{eq:tsj_org}
\end{equation}
where $Z_{\sigma_{v,s_j} \neq \boldsymbol{0}} = \sum_{\sigma_{v,s_j} \neq \boldsymbol{0}}P(\sigma_{v,s_j} \mid x_c,x_v,x_d)$ and $\sigma_{v,s_j}=\{\sigma_{v_0,s_j},\ldots,\sigma_{v_{N-1},s_j}\}$. However, exact computation of $t_{s_j}$ is difficult, as the selection probability $P(\sigma_{v,s_j} \mid x_c,x_v,x_d)$ depends on this optimization formulation and solver characteristics. Thus, we decompose the conditional probability applying Bayesian factorization:
\begin{equation}
\begin{split}
P(&\sigma_{v,s_j} \mid x_c,x_v,x_d) \\
&= P(n_{s_j},\sigma_{v,s_j} \mid x_c,x_v,x_d) \\
&= P(n_{s_j} \mid x_c,x_v,x_d)P(\sigma_{v,s_j} \mid x_c,x_v,x_d,n_{s_j}),
\end{split}
\end{equation}
where $n_{s_j}=\sum_{\sigma_{v_i,s_j} \in \sigma_{v,s_j}}\sigma_{v_i,s_j}$ denotes the number of vehicles assigned to station $s_j$.
Since customer assignment is prioritized, only $N-n_c$ vehicles are available
for station allocation. We approximate the first factor as a binomial distribution by ignoring the specific positions of customers and vehicles:
\begin{equation}
P(n_{s_j} \mid x_c,x_v,x_d) \approx B(N-n_c,1/n_s).
\end{equation}
In this approximation, we consider that the primary impact of customer demand is determining the total volume of surplus vehicles assigned to stations. Furthermore, while the specific vehicle positions $x_v$ and $x_d$ are critical for determining which specific vehicles are dispatched, this spatial dependency is explicitly modeled in the second factor. Therefore, in this factor, we focus solely on the macroscopic fleet size and ignore specific spatial coordinates, assuming a uniform distribution for the number of vehicles allocated to each station.
For the second factor, by ignoring the customer positions $x_c$ whose macroscopic effect is captured in the first factor, we approximate:
\begin{equation}
\begin{split}
P(\sigma_{v,s_j} \mid x_c,&x_v,x_d,n_{s_j}) \\
\approx &P(\sigma_{v,s_j} \mid x_v,x_d,n_{s_j}) \\
\propto &p(x_v,x_d \mid \sigma_{v,s_j},n_{s_j})P(\sigma_{v,s_j} \mid n_{s_j}),\label{eq:second}
\end{split}
\end{equation}
where the prior term $P(\sigma_{v,s_j} \mid n_{s_j})$ is assumed to be uniform, as no specific vehicles are prioritized for station allocation. 
By definition of conditional probability, the likelihood term is expressed as:
\begin{equation}
p(x_v,x_d \mid \sigma_{v,s_j},n_{s_j})=
\frac{p(x_v,x_d \mid \sigma_{v,s_j})}{\sum_{\sum{\sigma_{v,s_j}=n_{s_j}}}p(x_v,x_d \mid \sigma_{v,s_j})},
\end{equation}
where $\sum_{\sum{\sigma_{v,s_j}=n_{s_j}}}$ denotes the sum over subsets satisfying $\sum\sigma_{v,s_j}=n_{s_j}$. 
Since vehicles with shorter travel times are more likely to be selected due to the minimization of Eq. (\ref{eq:hb0}), we model the probability using an exponential distribution, similar to Eq. (\ref{eq:pxc}). Specifically, we define the probability based on the average travel time of the assigned vehicles:
\begin{equation}
p(x_v,x_d \mid \sigma_{v,s_j})=
\frac{1}{\theta_s}e^{-\frac{1}{\theta_s n_{s_j}}\sum_{\sigma_{v_i,s_j}=1}t(x_{v_i},x_{d_i},x_{s_j})},
\end{equation}
where $\theta_s$ is the average dispatch time of vehicles to stations.

However, $t_{s_j}$ depends on the current vehicle positions and the computation cost increases exponentially with $2^{(N-n_c)}$.
Therefore, we introduce the spatially averaged value $t_{s_j}^{\mathrm{avg}} = {\mathbb{E}}_{x_c,x_v,x_d}[t_{s_j}]$ and approximate it as follows (see Appendix for the detailed derivation):
\begin{align}
t_{s_j}^{\mathrm{avg}} &\approx {\mathbb{E}}_{x_{v_i},x_{d_i}}\left[t(x_{v_i},x_{d_i},x_{s_j})\right] \nonumber \\
&= \int_{x_{v_i},x_{d_i}}p(x_{v_i},x_{d_i})t(x_{v_i},x_{d_i},x_{s_j})dx_{v_i}dx_{d_i}. \nonumber \\
\intertext{Here, we decompose the integral into two components: occupied vehicles ($x_{v_i} \neq x_{d_i}$) and vacant vehicles ($x_{v_i} = x_{d_i}$).}
&= \int_{x_{v_i} \neq x_{d_i}}p(x_{v_i},x_{d_i})t(x_{v_i},x_{d_i},x_{s_j})dx_{v_i}dx_{d_i}+\int_{x_{v_i}}p(x_{v_i},x_{v_i} = x_{d_i})t(x_{v_i},x_{s_j})dx_{v_i} \nonumber \\
&= o\int_{x_{v_i} \neq x_{d_i}}p(x_{v_i},x_{d_i} \mid x_{v_i} \neq x_{d_i})
t(x_{v_i},x_{d_i},x_{s_j})dx_{v_i}dx_{d_i}+(1-o)\int_{x_{v_i}}p(x_{v_i} \mid x_{v_i} = x_{d_i})
t(x_{v_i},x_{s_j})dx_{v_i}.
\label{eq:tsj_avg_appx}
\end{align}
where $o=P(x_{v_i} \neq x_{d_i})$ represents the occupancy rate of vehicles.
Although the second integral can be estimated by analyzing the spatial distribution of vacant vehicles, we are concerned about its convergence time.
Therefore, instead of evaluating the integral directly, we use $\theta_s^{\mathrm{vac}}$ which denotes the average dispatch time of vacant vehicles to stations and is empirically observed during operation.
Finally, Eq. (\ref{eq:tsj_avg_appx}) is reformulated by decomposing the expectation value into occupied and vacant states (see Appendix for the detailed derivation):
\begin{equation}
\begin{split}
t_{s_j}^{\mathrm{avg}}&= o\int_{x_{c_i},x_{d_i}}\left\{p(x_{c_i},x_{d_i} \mid x_{v_i} \neq x_{d_i})\frac{t(x_{c_i},x_{d_i})}{2}+t(x_{d_i},x_{s_j})\right\}dx_{c_i}dx_{d_i}+(1-o)\theta_s^{\mathrm{vac}}.\label{eq:tsj_avg_conclusion}
\end{split}
\end{equation}
In the experiments presented in this paper, we assume independence between customer origins and destinations, approximating the joint distribution as $p(x_{c_i},x_{d_i} \mid x_{v_i} \neq x_{d_i}) \approx p(x_{c_i})p(x_{d_i}\mid x_{v_i} \neq x_{d_i})$. However, in actual operations, this probability can be explicitly estimated from historical data if significant correlations exist and sufficient data is available.

The customer appearance frequency is reflected in $p(x_{c_k} \mid s_j)$ of Eq. (\ref{eq:pxc}), leading stations closer to frequent customer areas to be prioritized.
Quantitatively, this behavior is consistent with practical expectations.
In addition, the destination frequency is considered in $t_{s_j}^{\mathrm{avg}}$ of Eq. (\ref{eq:tsj_avg_conclusion}), prioritizing stations farther from frequent destinations.
This prioritization is reasonable, as vehicles tend to accumulate around areas with frequent destinations.
On the other hand, $t_{s_j}$ of Eq. (\ref{eq:tsj_org}) incorporates the current vehicle destinations rather than statistical destination data.

\section*{Results}
\subsection*{Performance evaluation of the proposed formulation}
\begin{figure}[t!]
\centering
\includegraphics[width=13.5cm]{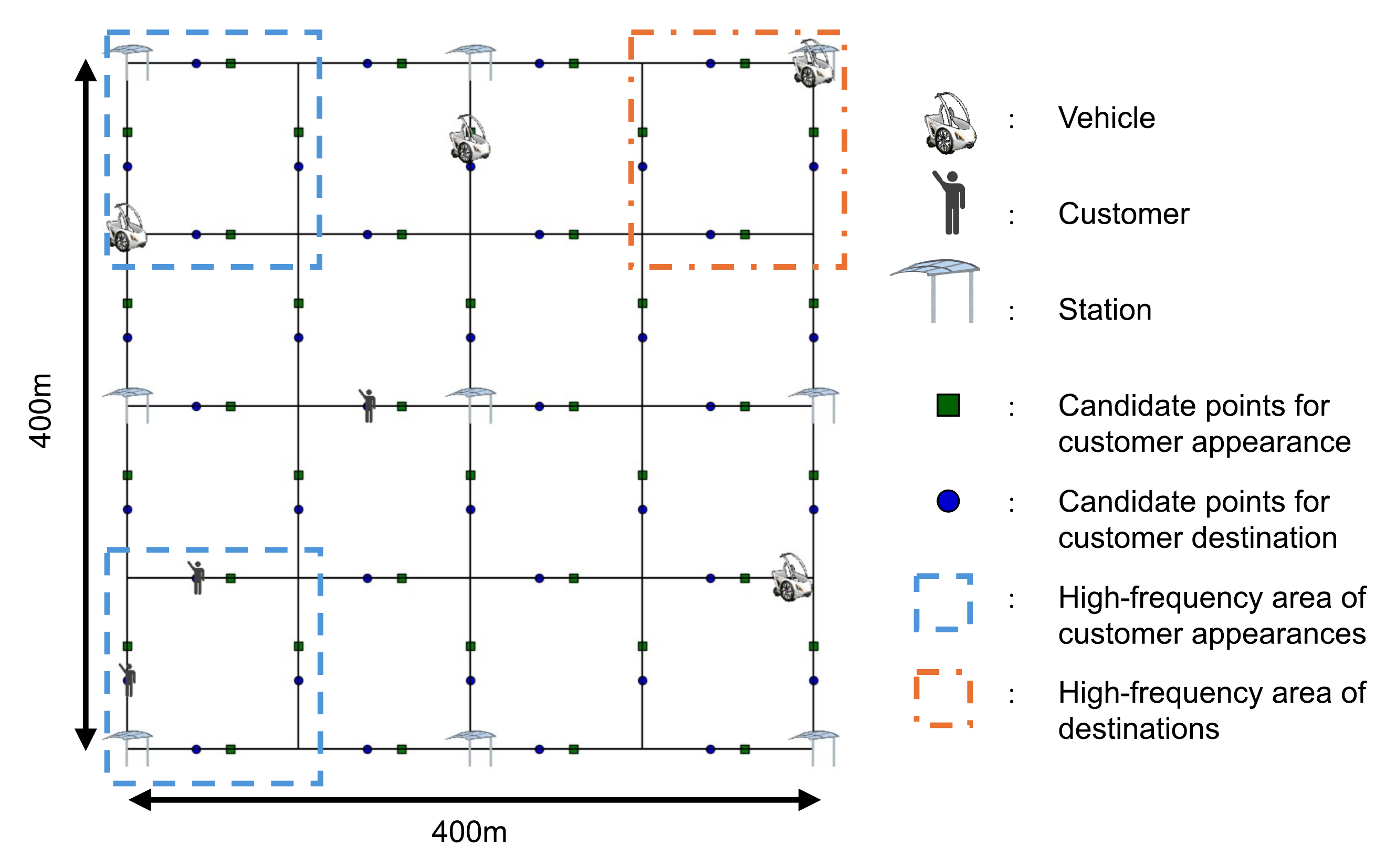}
\caption{Simulation environment for validating the proposed vehicle dispatch problem. The map consists of a $4 \times 4$ grid with six vehicles. The vehicle speed is $4~\mathrm{m/s}$. The average customer appearance interval is 50 seconds. High-frequency areas for both appearances and destinations have a 10-fold higher probability than other areas.}
\label{fig:sim}
\end{figure}

We developed a simulation framework to evaluate the proposed method, as depicted in Fig. \ref{fig:sim}.
The simulation assumes constant vehicle speed and models customer requests using a Poisson distribution.
To assess performance, we compared both the dynamic and static variants of the proposed method with a Greedy algorithm and a VRP-based approach formulated as a QUBO model tailored to the MMDP setting (see Appendix for formulation details).
We also performed an ablation study by removing $H_{A_1}$, which is a key component of our formulation.
The Greedy baseline prioritizes the customers with the longest waiting times (limited to the number of vehicles) and assigns them to vehicles based on the shortest travel time.
Surplus vehicles are directed to the nearest station.
The VRP-based approach utilizes a two-step look-ahead, enabling the schedule of two consecutive customer visits or a customer visit followed by a station return. To prevent an excessive increase in the number of variables, we limited the look-ahead horizon of the VRP. Note that even with this restriction, the number of variables remains approximately 1.4 times that of our proposed scheme.
Regarding the proposed method, the dynamic approach utilizes $t_{s_j}$ in Eq. (\ref{eq:tsj_org}) based on real-time vehicle positions.
In contrast, the static approach employs $t_{s_j}^\mathrm{avg}$ in Eq. (\ref{eq:tsj_avg_conclusion}), which is calculated by averaging the vehicle distribution using historical data.
As an initial validation step, the schemes except for Greedy use Simulated Quantum Annealing (SQA) as a substitute for real quantum hardware. Since annealing performance is generally sensitive to the problem size, this setup also enables us to observe the potential performance degradation caused by the larger number of variables.

\begin{figure}[t!]
\centering
\includegraphics[width=17cm]{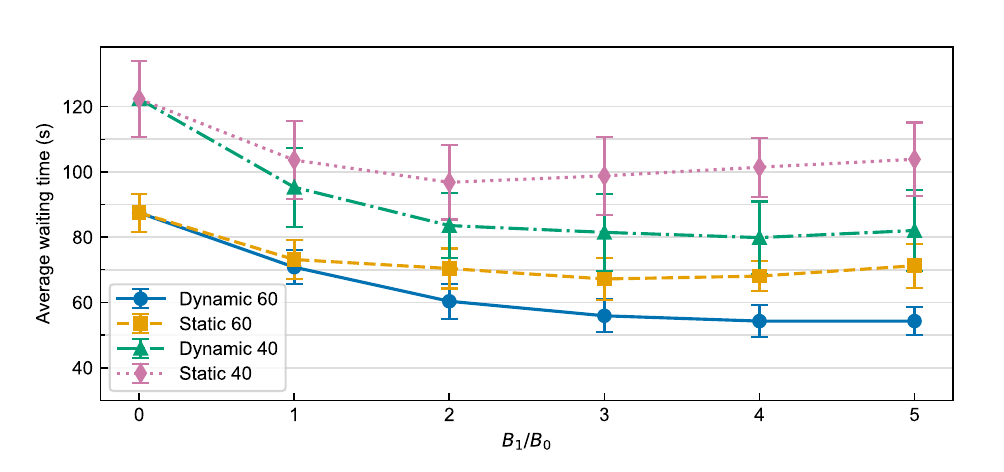}
\caption{Average waiting time of the customers according to the weight $B_1$ compared to $B_0$. We measured 10 trials, and each trial lasted for 10,000 seconds.}
\label{fig:b1}
\end{figure}

First, we evaluated the impact of the weight ratio $B_1/B_0$ on the average waiting time, with $B_0$ in Eq. (\ref{eq:total}) fixed at 0.1.
The results are shown in Fig. \ref{fig:b1}.
Calibrating these parameters is crucial as they balance the trade-off between assigning distant stations and minimizing immediate travel time.
Thus, we determined these values empirically.
For both the dynamic and static approaches, the optimal $B_1$ trend for minimizing the waiting time was consistent, independent of customer arrival frequency.
Based on these results, we adopted $B_1 = 0.3$ and $B_0 = 0.1$ for this study.

\begin{table}[t!]
\centering
(a) Low-frequency customer requests (60 s/request).
\begin{tabular}{lccccc}
\hline
Metric & Greedy & VRP & \makecell{Proposed\\w/o $H_{A_1}$} & \makecell{Proposed\\(Static)} & \makecell{Proposed\\(Dynamic)} \\
\hline
Wait time (s) & 81.4 $\pm$ 5.0 & 102.6 $\pm$ 5.9 & 87.5 $\pm$ 5.9 & 67.2 $\pm$ 6.3 & \textbf{55.6} $\pm$ 5.0 \\
Num. of waiting customers & 1.3 $\pm$ 0.3 & 1.7 $\pm$ 0.2 & 1.5 $\pm$ 0.2 & 1.1 $\pm$ 0.2 & \textbf{0.9} $\pm$ 0.1 \\
Customer dispatch time (s) & 82.5 $\pm$ 5.2 & 82.3 $\pm$ 3.4 & 80.2 $\pm$ 4.3 & 67.2 $\pm$ 5.6 & \textbf{58.4} $\pm$ 5.2 \\
Station dispatch time (s) & 76.8 $\pm$ 4.6 & 66.6 $\pm$ 4.3 & \textbf{65.9} $\pm$ 1.7 & 73.4 $\pm$ 2.8 & 74.2 $\pm$ 3.6 \\
Inter-vehicle distance (m) & 355.8 $\pm$ 10.5 & 375.5 $\pm$ 8.0 & 367.5 $\pm$ 8.4 & 412.6 $\pm$ 5.2 & \textbf{417.6} $\pm$ 7.6 \\
Total travel distance (km) & \textbf{14.7} $\pm$ 1.2 & 27.2 $\pm$ 1.1 & 18.0 $\pm$ 0.8 & 16.5 $\pm$ 0.9 & 19.3 $\pm$ 1.2 \\
\hline
\end{tabular}
\vspace{2mm}

(b) High-frequency customer requests (40 s/request)
\begin{tabular}{lccccc}
\hline
Metric & Greedy & VRP & \makecell{Proposed\\w/o $H_{A_1}$} & \makecell{Proposed\\(Static)} & \makecell{Proposed\\(Dynamic)} \\
\hline
Wait time (s) & 108.9 $\pm$ 14.1 & 182.7 $\pm$ 44.0 & 122.4 $\pm$ 11.6 & 98.8 $\pm$ 12.0 & \textbf{81.5} $\pm$ 11.7 \\
Num. of waiting customers & 2.8 $\pm$ 0.5 & 4.7 $\pm$ 1.3 & 3.1 $\pm$ 0.5 & 2.5 $\pm$ 0.4 & \textbf{2.1} $\pm$ 0.4 \\
Customer dispatch time (s) & 103.5 $\pm$ 9.0 & 98.7 $\pm$ 5.3 & 94.9 $\pm$ 4.8 & 87.6 $\pm$ 5.2 & \textbf{75.3} $\pm$ 5.7 \\
Station dispatch time (s) & 91.7 $\pm$ 3.8 & 83.8 $\pm$ 8.6 & \textbf{74.0} $\pm$ 2.2 & 93.8 $\pm$ 5.8 & 91.5 $\pm$ 5.4 \\
Inter-vehicle distance (m) & 394.6 $\pm$ 6.3 & 390.5 $\pm$ 6.0 & 395.7 $\pm$ 7.4 & \textbf{430.9} $\pm$ 4.7 & 429.6 $\pm$ 5.3 \\
Total travel distance (km) & \textbf{21.5} $\pm$ 1.5 & 31.5 $\pm$ 0.9 & 22.9 $\pm$ 1.0 & 21.9 $\pm$ 1.0
& 25.5 $\pm$ 1.0 \\
\hline
\end{tabular}
\caption{Performance comparison. Highlighted values indicate the minimum metrics, except for the inter-vehicle distance, where larger values imply better vehicle dispersion (avoiding clustering). Dispatch time is measured at the moment of vehicle assignment. Since assignments may change due to dynamic on-demand requests, the actual customer waiting time is longer than the dispatch time.}
\label{tbl:comparison}
\end{table}

Next, Table $\ref{tbl:comparison}$ presents the performance of each scheme under low and high request frequency scenarios. The proposed dynamic approach outperforms other methods in service quality metrics, such as customer waiting time, in both scenarios. Conversely, it exhibits the longest station dispatch time, indicating that the scheme directs vehicles to appropriate stations even if they are distant. This scheme also tends to avoid assigning stations located near other vehicles. Because real-time vehicle positions are considered, the optimal station assignment changes frequently. This causes vehicles to move between stations often, thereby increasing the total travel time.
The static approach exhibits similar characteristics to the dynamic one except for the travel time.
The term $t_{s_j}^{\mathrm{avg}}$ effectively guides vehicles to high-frequency request areas without relying on real-time vehicle positions.
Furthermore, since the static approach does not account for real-time positions, it prevents unnecessary station-to-station travel. Both proposed schemes maintain long inter-vehicle distances, indicating effective spatial dispersion of vehicles. In the low-demand scenario, the dynamic approach achieves the highest dispersion by leveraging real-time position data. Under the high-frequency scenario, both schemes yield similar values. We attribute this to the difficulty of maintaining dispersion when the vehicle operation rate is high.
The Greedy approach and the ablation model without $H_{B_1}$, which prioritize minimizing immediate travel time, show similar results. Comparing these against the proposed method demonstrates that $H_{B_1}$ significantly improves service quality. The VRP method does not show superior performance despite its two-step look-ahead capability. In the MMDP setting, the problem state changes frequently due to real-time requests, rendering long-term multi-step planning ineffective.
In low-frequency scenarios, the second time slots in the VRP formulation are rarely utilized. Consequently, the VRP logic effectively degenerates into a single-step distance minimization, behaving similarly to the proposed scheme without the $H_{A_1}$ term.
However, the VRP performance is notably inferior. This degradation is attributed to the larger number of variables, which hinders the solver's convergence and leads to unstable solutions.
Interestingly, although the planned travel time (dispatch time) of the VRP is shorter than that of the Greedy approach, the actual customer waiting time is longer.
Furthermore, the VRP scheme suffers from frequent solution oscillations, where assignments flip between slots. Moreover, customers in low-density regions are often continuously postponed to the second slot. These behaviors cause certain customers to wait for excessively long periods, making the average performance worse than in high-frequency scenarios. Although these phenomena could theoretically be mitigated by introducing weights for long-waiting customers, the results from the low-frequency scenario indicate that the benefits of multi-step planning do not compensate for the performance penalty caused by the increased number of variables.

\begin{figure}[t!]
\centering
\includegraphics[width=16.5cm]{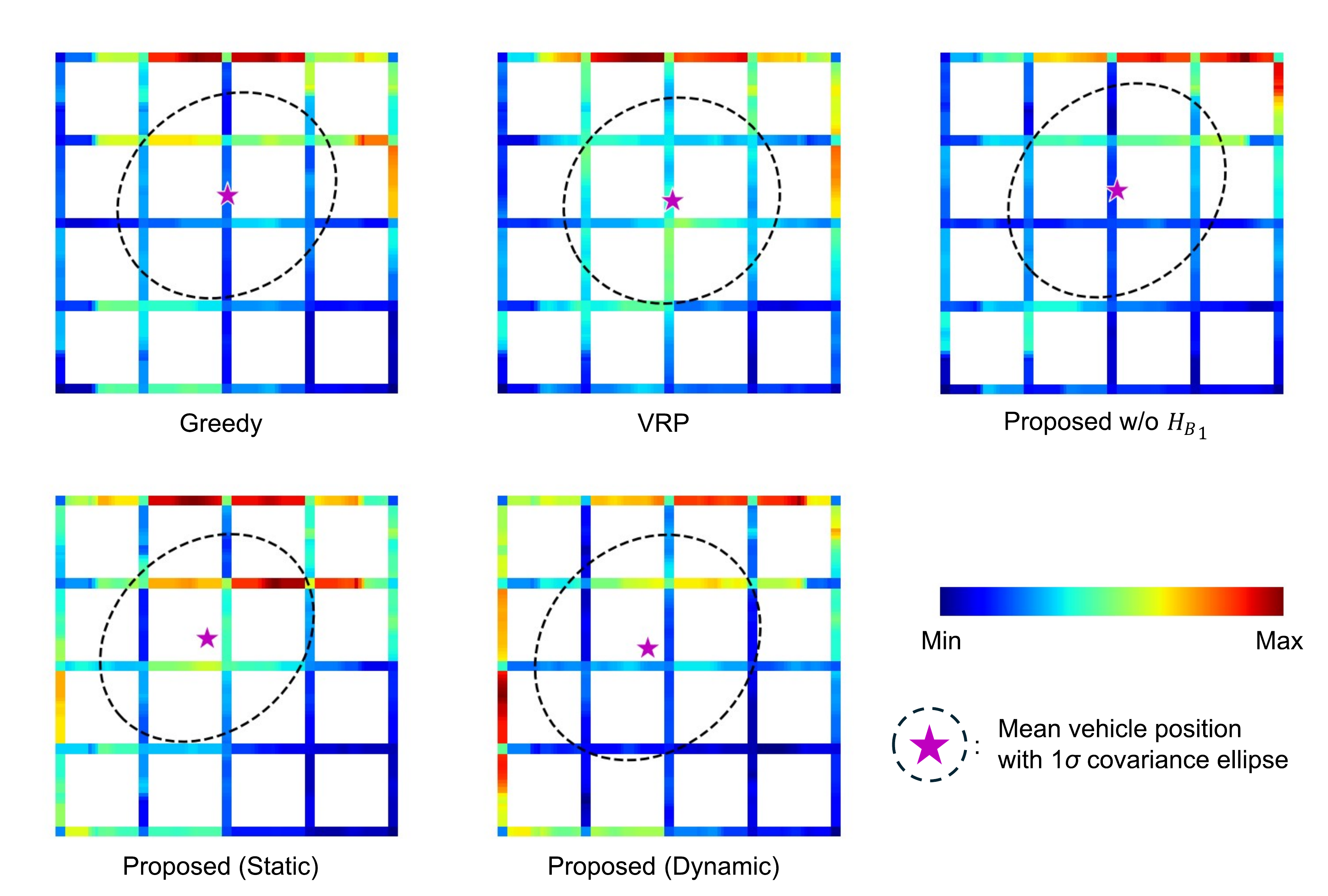}
\caption{Comparison of vehicle position statistics under different dispatch strategies.
Each panel shows a heatmap of vacant vehicle positions along with the fitted Gaussian distribution. The customer request rate is high $40~\mathrm{s/request}$.}
\label{fig:veh_pos}
\end{figure}
To demonstrate that our schemes effectively guide vehicles to high-demand areas, Fig. \ref{fig:veh_pos} illustrates the vehicle distribution. While the baselines show little difference in vehicle position distribution, the proposed schemes exhibit distinct patterns. The proposed schemes incorporating $H_{B_1}$ encourage vehicles to move from high-destination areas to high-request areas. The dynamic approach is particularly responsive. The static scheme relies on the destination distribution, typically selecting stations far from the top-right area. However, as a general trend, many vehicles route through the top area to return to stations after service. The dynamic approach actively utilizes
stations in the bottom area, because vehicles concentrated in the top area exert a repulsive force on station allocation.

\subsection*{Characteristics of QA for the proposed QUBO}
\begin{figure}[t!]
\centering
\includegraphics[width=17cm]{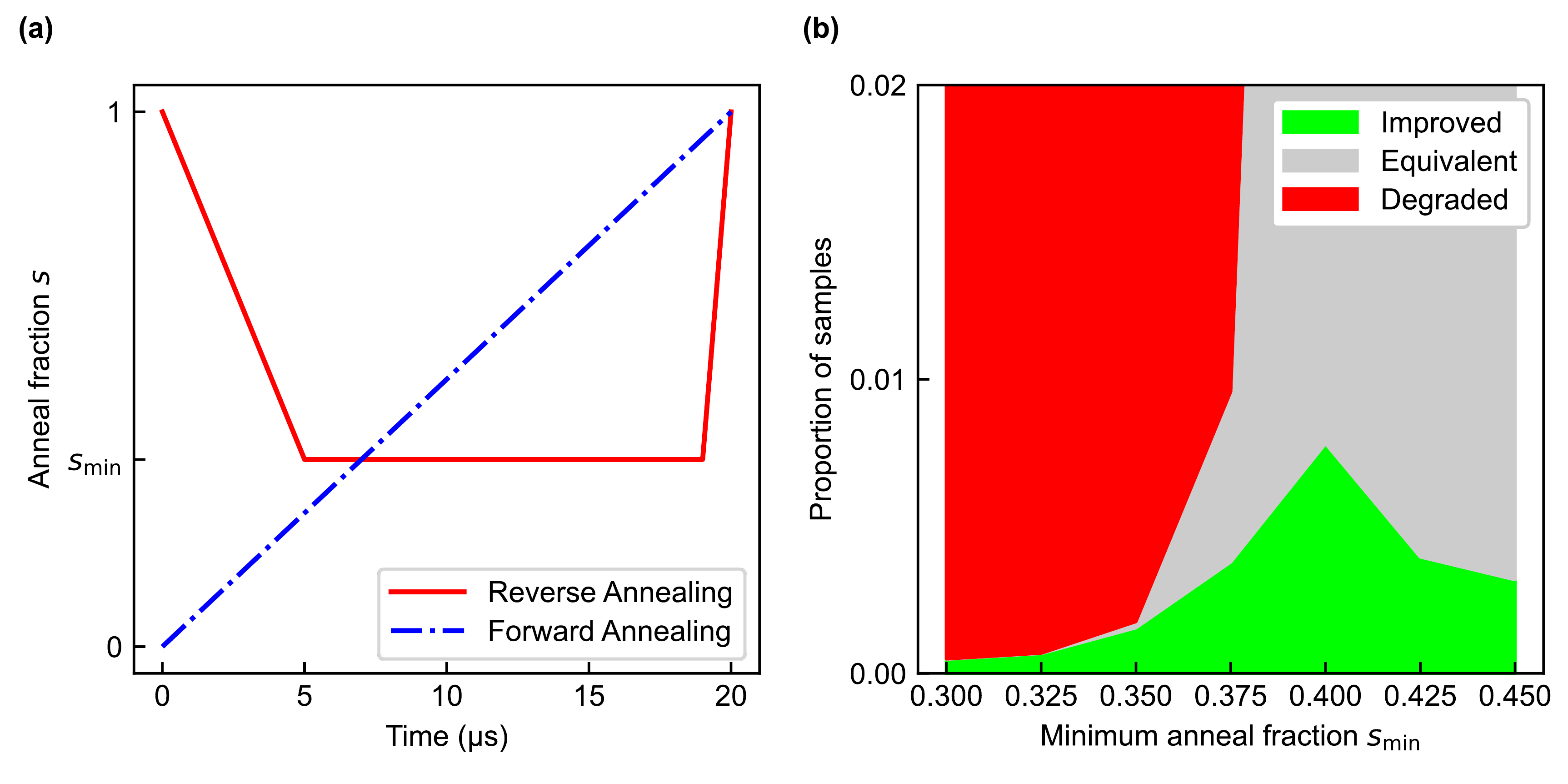}
\caption{Parameter tuning of the minimum anneal fraction $s_{\mathrm{min}}$. We generated 30 random instances using the simulator and solved them using the proposed static formulation.
(a) Annealing schedules. We adopted the default D-Wave FA schedule as a baseline. The RA schedule follows Ref.~\cite{haba2022travel}, where the pause point $s_{\mathrm{min}}$ is a tunable parameter. The total annealing time for RA is set to match the FA duration.
(b) Distribution of solution quality relative to initial states as a function of $s_{\mathrm{min}}$.}
\label{fig:s_min}
\end{figure}
We evaluated actual quantum hardware using the D-Wave Advantage 1.1 system applied to our problem setting and formulation. Each problem run returned 10,000 samples. For comparison and obtaining exact solutions, we employed Gurobi Optimizer 13.0 running on a computer equipped with an AMD Ryzen 9 7940HS and 64 GB RAM. Gurobi is widely recognized as a standard and high-performance solver for integer programming problems.
RA initiates the process from a classical initial state. Our strategy extends the greedy heuristic by incorporating a priority-based station assignment rule. After assigning vehicles to customers, we assign the nearest available vehicle to the station with the highest $\tau_{s_j}$, decrement the value by 1, and repeat this cycle until no surplus vehicles remain.
We employed two metrics to evaluate performance.
Time-to-solution (TTS), a metric for optimizer performance, is defined as
\begin{equation}
\mathrm{TTS}(p) = t_c\frac{\log{\left(1-p\right)}}{\log{\left(1-p_\mathrm{opt}\right)}}
\end{equation}
where $p$ represents the target cumulative success probability (typically 0.99), $p_\mathrm{opt}$ is the probability of obtaining the optimal solution in a single trial, and $t_c$ is the computation time per trial. Here, $\mathrm{TTS}(p)$ denotes the computation time required to find the exact solution with probability $p$.
The relative minimum residual energy, which quantifies solution quality, is defined as
\begin{equation}
E_\mathrm{res} = \frac{E_{\mathrm{min}} - E_\mathrm{opt}}{E_\mathrm{opt}}
\end{equation}
where $E_{\mathrm{min}}$ is the minimum energy among the obtained samples, and $E_\mathrm{opt}$ is the ground state energy.

Figure \ref{fig:s_min} illustrates the annealing fraction profile, highlighting the tunable parameter $s_{\mathrm{min}}$, and the relationship between solution improvement and $s_{\mathrm{min}}$. The annealing fraction represents the progress of the annealing. In Forward Annealing (FA), it transitions linearly from zero to one. In contrast, for RA, it starts at one, gradually decreases to $s_{\mathrm{min}}$, pauses, and finally returns to one. A lower $s_{\mathrm{min}}$ encourages exploration, but may result in solutions drifting further from the initial state. Based on these results, we selected $s_{\mathrm{min}}=0.4$ as this setting yielded the highest number of solutions surpassing the initial state.

\begin{figure}[t!]
\centering
\includegraphics[width=17cm]{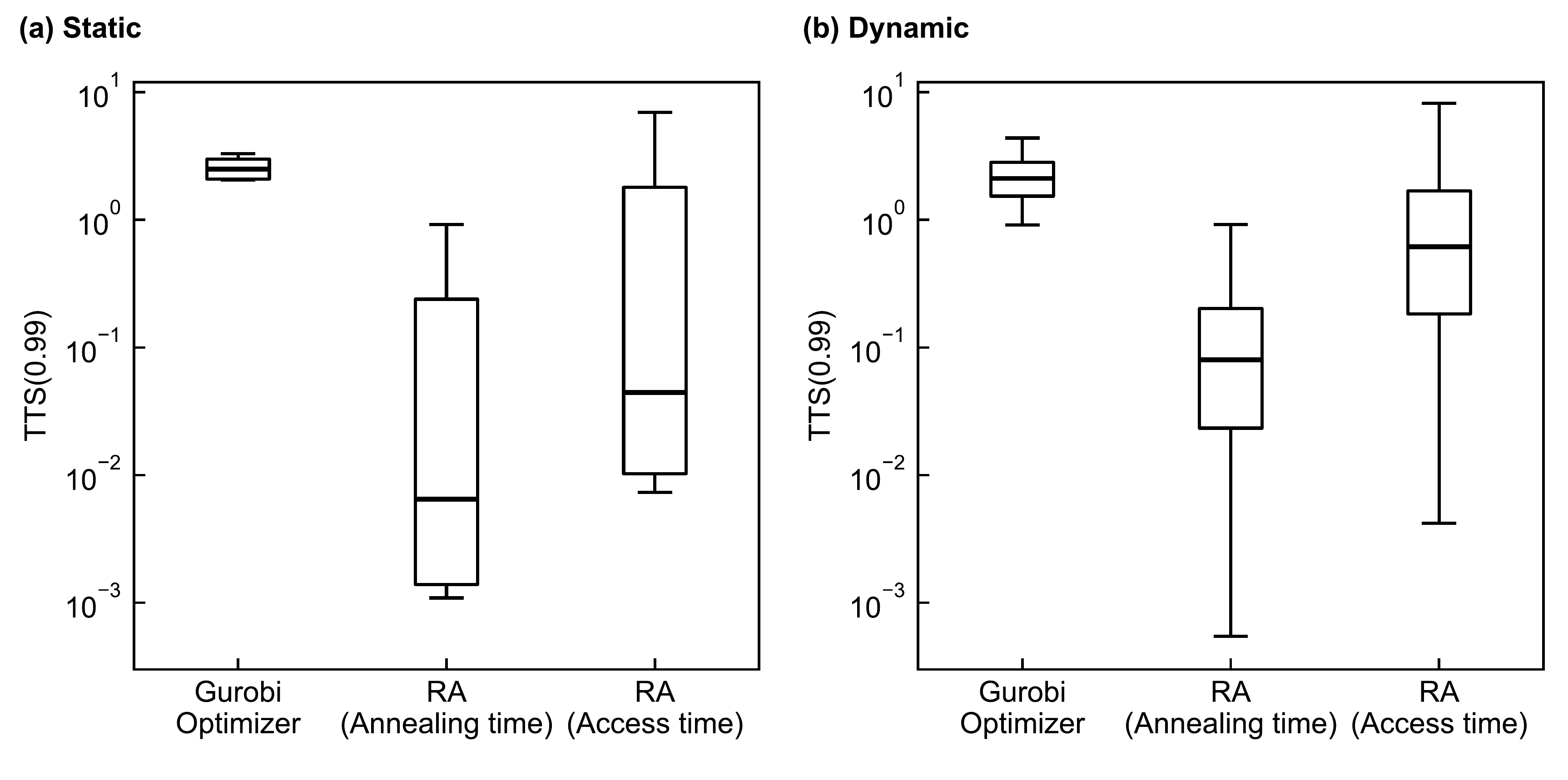}
\caption{Comparison of TTS between Gurobi Optimizer and RA. The results include only those instances that were exactly solved by RA. RA attempted 30 problems, consistent with the previous experiment.
(a) Static formulation. Four problems were solved. 
(b) Dynamic formulation. Six problems were solved.}
\label{fig:tts}
\end{figure}

Figure \ref{fig:tts} compares the performance of the optimizers for both the proposed static and dynamic formulations.
The annealing time per trial was set to 20 {\textmu}s, as shown in Fig. \ref{fig:s_min}. Although the pure annealing time is very small, the QPU access time, which encompasses programming, annealing, readout, and delays, is significantly longer. While the QPU access time may decrease with future technological advancements, it is currently non-negligible. Therefore, we present two types of TTS: one based on the pure annealing time and another on the total QPU access time. Gurobi exhibited less variation in performance compared to RA. TTS of RA varied significantly depending on problem parameters, though it outperformed Gurobi in certain cases, even when considering the full QPU access time. The number of solved instances was higher for the dynamic formulation than for the static one. The average residual energies of the initial states for the static and dynamic formulations were 0.088 $\pm$ 0.054 and 0.056 $\pm$ 0.036, respectively. The difference implies that the dynamic formulation is easier to solve, a result consistent with the findings in Ref.~\cite{haba2022travel}. In the static formulation, averaging vehicle positions reduces the variance of $\tau_{s_j}$ among stations compared to the dynamic case. Consequently, the initial state generation strategy is more effective at selecting the correct stations in the dynamic approach.

\begin{figure}[t!]
\centering
\includegraphics[width=17cm]{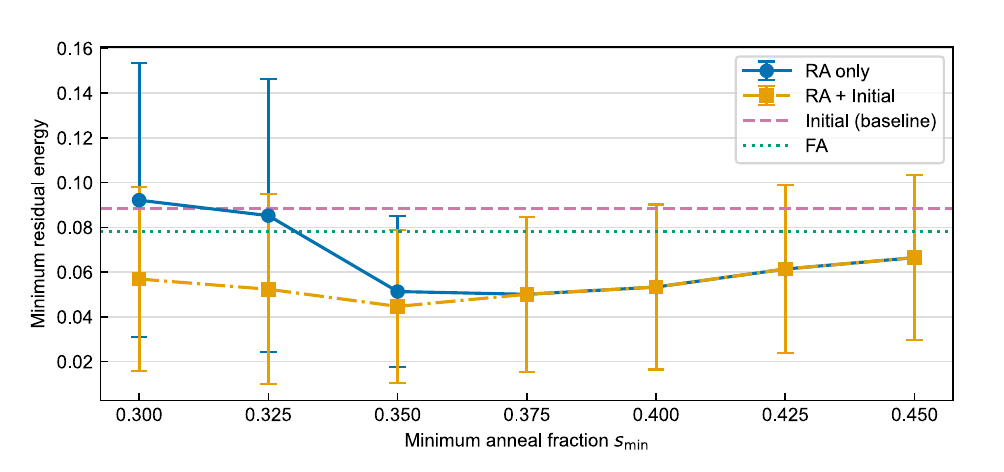}
\caption{Residual energies obtained by RA using the static formulation. The problem settings are identical to those in the previous experiments. The average residual energy of the initial states serves as the baseline, and the forward annealing result is provided for reference.}
\label{fig:res_eng}
\end{figure}
However, RA failed to find the exact optimal solution for most problems. Practically, micro-mobility dispatch operations do not require a strictly optimal solution, provided the operation is sufficiently efficient. Therefore, to assess the potential for obtaining better solutions, Fig. \ref{fig:res_eng} depicts the relationship between the relative minimum residual energy and the annealing parameter $s_{\mathrm{min}}$. The results demonstrate that when $s_{\mathrm{min}}$ is tuned to appropriate values, RA outperforms both the initial states and FA. We also consider a simple scheme that includes the initial state in the final solution set to prevent quality degradation at low $s_{\mathrm{min}}$ values, where exploration is aggressive. The residual energy was minimized at $s_{\mathrm{min}}=0.35$ for RA when including initial states. This indicates that aggressive exploration improves solution quality when safeguarded by the inclusion of the initial state, potentially enhancing operational efficiency.

\section*{Discussion}
This study proposed a novel QUBO formulation for the micro-mobility dispatch problem, incorporating historical data via a Bayesian approach to estimate the ideal distribution of waiting vehicles. We presented two incorporation schemes: a dynamic approach utilizing real-time vehicle positions and a static approach based solely on statistical data. Both approaches outperformed the baseline algorithms—the Greedy heuristic and the VRP-based formulation—in terms of customer service metrics. Notably, the dynamic approach achieved the most efficient operation, minimizing customer waiting times. However, this improvement came at the cost of increased total travel time, resulting from the active repositioning of vehicles based on real-time states. In contrast, the static approach demonstrated a balanced improvement in service quality without a significant increase in travel distance. In the evaluation using actual quantum hardware, RA outperformed the Gurobi Optimizer in terms of TTS under specific problem parameters. Furthermore, our results revealed that RA achieves lower residual energy compared to FA. We also demonstrated that including the initial state in the final solution set effectively prevents quality degradation. This strategy enables the adoption of a lower minimum annealing fraction, thereby encouraging more aggressive exploration of the solution space, safeguarded by the initial state.

The increase in total travel time, which directly correlates with energy consumption, is a critical factor alongside operational efficiency. Improving the service quality of the static approach further without increasing travel time presents a challenge, as the dynamic approach inherently achieves optimal vacant vehicle positions by adaptively changing station allocations based on current vehicle locations. Therefore, for practical implementation, a hybrid scheme combining the dynamic and static approaches should be considered. By introducing a control parameter to blend these schemes, it would be possible to manage the trade-off between service quality (waiting time) and operational cost (energy consumption) according to system requirements.
Furthermore, the proposed formulation involves operational parameters, such as average dispatch time, which are embedded within the QUBO coefficients. In continuous operation, these parameters and the resulting performance metrics affect each other cyclically. Future work must examine the stability of this feedback loop to ensure robust long-term operation. Additionally, our formulation relies on several approximations regarding probability distributions. Validating these approximations and exploring model-free estimation methods, such as neural networks, could further enhance the accuracy and performance of the dispatch logic.
For instance, the number of vehicles assigned to a station is currently approximated as a Binomial distribution, assuming the stations are selected uniformly and independently. However, an increase in the number of vehicles may introduce biases between stations, which could potentially degrade the dispatch performance.

To evaluate the feasibility of the proposed method in real-world scenarios, investigations utilizing actual urban road networks, large-scale demand data, and larger fleet sizes are necessary. For large-scale operations that exceed the qubit capacity of current hardware, problem decomposition schemes will be essential. From an algorithmic perspective, overcoming hardware constraints remains a key challenge. The constraint relaxation methods proposed by Ohzeki~\cite{ohzeki2020breaking} can convert equality constraints into linear biases, potentially conserving qubit resources. Additionally, modeling the maximum capacity of stations requires inequality constraints. The relaxation scheme for inequality constraints~\cite{takabayashi2025subgradient} could be valuable for integrating such realistic limitations into our formulation. Furthermore, the clustering-based problem decomposition~\cite{quang2025quantum} has proven effective for handling large-scale variables. In our context, our experiments indicate that the diversity of $\tau_{s_j}$ facilitates RA in reaching optimal solutions. This finding suggests that $\tau_{s_j}$ characteristics could serve as critical criteria for defining efficient clustering groups in MMDP. Finally, regarding RA, this study utilized a Greedy-based initial solution. However, analogous to Model Predictive Control (MPC), where previous solutions serve as a warm-start, utilizing the solution from the preceding time step could provide a superior initial state. Although our problem setting assumes event-driven operations, the solution from the previous time step likely retains beneficial information for the current step. Investigating this warm-start strategy for RA is a promising direction for real-time applications. Moreover, it is crucial to elucidate the relationship between residual energy and operational metrics within a realistic simulation environment, and to explore strategies for reducing residual energy to facilitate near-term practical application.

\section*{Appendix}
\subsection*{Supplementary formulation}
We provide a detailed derivation of Eq. (\ref{eq:tsj_org}) and Eq. (\ref{eq:tsj_avg_appx}).
Starting from the definition of $t_{s_j}$ in Eq. (\ref{eq:tsj_org}), the expected average travel time is expressed as:
\begin{align}
{\mathbb{E}}_{x_{c},x_{v},x_{d}}\left[t_{s_j}^{\mathrm{avg}}\right] =& \int_{x_c,x_v,x_d}p(x_c,x_v,x_d)\sum_{\sigma_{v,s_j} \neq \boldsymbol{0}}\frac{P(\sigma_{v,{s_j}} \mid x_c, x_v, x_d)}{Z_{\sigma_{v,s_j} \neq \boldsymbol{0}}}
\frac{\sum_{\sigma_{v_i,s_j} \in \sigma_{v,s_j}}\sigma_{v_i,s_j}t(x_{v_i},x_{d_i},x_{s_j})}{\sum_{\sigma_{v_i,s_j} \in \sigma_{v,s_j}}\sigma_{v_i,s_j}}dx_{c}dx_{v}dx_{d} \nonumber \\
=& \sum_{\sigma_{v,s_j} \neq \boldsymbol{0}}\int_{x_c,x_v,x_d}p(\sigma_{v,{s_j}}, x_c, x_v, x_d \mid \sigma_{v,s_j} \neq \boldsymbol{0})dx_c
\frac{\sum_{\sigma_{v_i,s_j} \in \sigma_{v,s_j}}\sigma_{v_i,s_j}t(x_{v_i},x_{d_i},x_{s_j})}{\sum_{\sigma_{v_i,s_j} \in \sigma_{v,s_j}}\sigma_{v_i,s_j}}dx_vdx_d \nonumber \\
=& \sum_{\sigma_{v,s_j} \neq \boldsymbol{0}}
\frac{\sum_{\sigma_{v_i,s_j} \in \sigma_{v,s_j}}\sigma_{v_i,s_j}\int_{x_v,x_d}p(\sigma_{v,{s_j}},x_v,x_d \mid \sigma_{v,s_j} \neq \boldsymbol{0})t(x_{v_i},x_{d_i},x_{s_j})dx_vdx_d}{\sum_{\sigma_{v_i,s_j} \in 
\sigma_{v,s_j}}\sigma_{v_i,s_j}}. \nonumber \\
\intertext{Since the vehicles are identical, the expectation of the average is equivalent to the expectation of a single vehicle. Thus the summation in the numerator cancels the denominator:}
=& 
\int_{x_v,x_d}\sum_{\sigma_{v,s_j} \neq \boldsymbol{0}}p(\sigma_{v,{s_j}},x_v,x_d \mid \sigma_{v,s_j} \neq \boldsymbol{0})t(x_{v_i},x_{d_i},x_{s_j})dx_vdx_d \nonumber \\
=& \int_{x_{v_i},x_{d_i}}p(x_{v_{i}},x_{d_{i}} \mid \sigma_{v,s_j} \neq \boldsymbol{0})t(x_{v_i},x_{d_i},x_{s_j})dx_{v_i}dx_{d_i} \approx {\mathbb{E}}_{x_{v_i},x_{d_i}}\left[t(x_{v_i},x_{d_i},x_{s_j})\right].
\end{align}
Here, we approximate $p(x_{v_{i}},x_{d_{i}} \mid \sigma_{v,s_j} \neq \boldsymbol{0})$ as $p(x_{v_{i}},x_{d_{i}})$. Since the proposed static scheme operates independently of real-time vehicle locations, the vehicle distribution is considered statistically independent of the allocation status.

Next, we approximate Eq. (\ref{eq:tsj_avg_appx}) to derive Eq. (\ref{eq:tsj_avg_conclusion}). The second integral term (vacant vehicles) is approximated by the observed average dispatch time $\theta_s^{\mathrm{vac}}$.
Thus,
\begin{equation}
t_{s_j}^{\mathrm{avg}} \approx o\int_{x_{v_i} \neq x_{d_i}}p(x_{v_i},x_{d_i} \mid x_{v_i} \neq x_{d_i})t(x_{v_i},x_{d_i},x_{s_j})dx_{v_i}dx_{d_i}+(1-o)\theta_s^{\mathrm{vac}}.\label{eq:tsj_avg_start}
\end{equation}
The first integral can be decomposed into the leg to the destination and the leg to the station:
\begin{multline}
\int_{x_{v_i} \neq x_{d_i}}p(x_{v_i},x_{d_i} \mid x_{v_i} \neq x_{d_i})t(x_{v_i},x_{d_i},x_{s_j})dx_{v_i}dx_{d_i}= \\
\int_{x_{d_i}}\int_{x_{v_i} \neq x_{d_i}}p(x_{v_i},x_{d_i} \mid x_{v_i} \neq x_{d_i})t(x_{v_i},x_{d_i})dx_{v_i}
 + p(x_{d_i} \mid x_{v_i} \neq x_{d_i})t(x_{d_i},x_{s_j})dx_{d_i}.
\end{multline}
We introduce the trip start location $x_{c_i}$ for vehicle $v_i$ into the first term in the integral.
\begin{equation}
\begin{split}
&\int_{x_{d_i}}\int_{x_{v_i} \neq x_{d_i}}p(x_{v_i},x_{d_i} \mid x_{v_i} \neq x_{d_i})t(x_{v_i},x_{d_i})dx_{v_i}dx_{d_i}\\
&\quad= \int_{x_{c_i},x_{d_i}}\int_{x_{c_i}\leq x_{v_i}< x_{d_i}}p(x_{c_i},x_{v_i},x_{d_i} \mid x_{v_i} \neq x_{d_i})t(x_{v_i},x_{d_i})dx_{v_i}dx_{c_i}dx_{d_i} \\
&\quad= \int_{x_{c_i}, x_{d_i}} p(x_{c_i},x_{d_i} \mid x_{v_i} \neq x_{d_i})\int_{x_{c_i}\leq x_{v_i}< x_{d_i}}p(x_{v_i} \mid x_{c_i}, x_{d_i}, x_{v_i} \neq x_{d_i})t(x_{v_i},x_{d_i})dx_{v_i}dx_{c_i}dx_{d_i}\label{eq:intro_xc}
\end{split}
\end{equation}
where the integral $\int_{x_{c_i}\leq x_{v_i}< x_{d_i}}$ denotes integration along the path from $x_{c_i}$ to $x_{d_i}$.
The probability of a vehicle existing in a certain path segment is proportional to the duration of stay in that area.
Thus, the probability density at $x_{v_i}$ can be expressed as
\begin{equation}
p(x_{v_i} \mid x_{c_i}, x_{d_i}, x_{v_i} \neq x_{d_i})dx_{v_i} = \frac{dt}{t(x_{c_i}, x_{d_i})},
\end{equation}
where $t$ represents the time elapsed from the departure at $x_{c_i}$. Note that $t(x_{v_i}, x_{d_i}) = t(x_{c_i}, x_{d_i}) - t$.
Substituting this into the integral over $x_{v_i}$ in Eq. (\ref{eq:intro_xc}):
\begin{equation}
\begin{split}
\int_{x_{c_i}\leq x_{v_i}< x_{d_i}}p(x_{v_i} \mid x_{c_i}, x_{d_i}, x_{v_i} \neq x_{d_i})t(x_{v_i},x_{d_i})dx_{v_i} 
&= \int_{0}^{t(x_{c_i},x_{d_i})} \frac{1}{t(x_{c_i}, x_{d_i})} (t(x_{c_i}, x_{d_i}) - t) dt \\
&= \frac{1}{t(x_{c_i}, x_{d_i})} \left[ t(x_{c_i}, x_{d_i})t - \frac{t^2}{2} \right]_{0}^{t(x_{c_i},x_{d_i})} \\
&= \frac{t(x_{c_i}, x_{d_i})}{2}. \label{eq:t/2}
\end{split}
\end{equation}
Finally, by substituting Eq. (\ref{eq:intro_xc}) into Eq. (\ref{eq:tsj_avg_start}), applying the result of Eq. (\ref{eq:t/2}), and changing the index of $x_{c_i}$ from $i$ to $k$ to correspond with the notation in the main text, we obtain Eq. (\ref{eq:tsj_avg_conclusion}).

\subsection*{Details of VRP-based QUBO formulation for MMDP}
We introduce two types of binary variables: $\sigma_{v,s} \in \{0,1\}^{Nn_s}$ and $\sigma_{v,c,t} \in \{0,1\}^{2Nn_c}$. The variable $\sigma_{v_i,s_j}$ indicates whether vehicle $v_i$ is dispatched to station $s_j$. For customer assignments, $\sigma_{v_i,c_k,t}$ represents whether customer $c_k$ is visited by vehicle $v_i$ at position $t$ in its service sequence, where $t=1$ and $t=2$ denote the first and second customer, respectively. Each vehicle can serve up to two customers in this VRP-based baseline. Therefore, if the number of customers exceeds $2N$, we extract the $2N$ earliest-received customers and set $n_c=2N$.
If $\sum_k\sigma_{v_i, c_k, 1}=1$, then $v_i$ serves exactly one first customer, and $\sigma_{v_i, s_j}=1$ indicates that the vehicle proceeds to station $s_j$ after its service. Conversely, if $\sum_k\sigma_{v_i,c_k,1}=0$, no customer is assigned to vehicle $v_i$ and $\sigma_{v_i,s_j}=1$ simply indicates that the vehicle moves directly to station $s_j$.

All customers must be assigned to exactly one vehicle, which is formulated as
\begin{equation}
H_{A_0} = \sum_k\left(1-\sum_{i, t}\sigma_{v_i,c_k,t}\right)^2.
\end{equation}
We also aim to maximize the number of waiting customers assigned to the first slot. We define the target number of first customers as $n_{\mathrm{1st}}=\min{\left(N, n_c\right)}$ and the constraint Hamiltonian to fill the first slots is expressed as
\begin{equation}
H_{A_1} = \left(n_{\mathrm{1st}}-\sum_{i,k}\sigma_{v_i,c_k,1}\right)^2.\label{eq:vrp_ha1}
\end{equation}
Each vehicle operates in one of three route patterns: (1) direct travel to a station, (2) travel to a station after serving the first customer, or (3) service of a second customer after the first. If a station is selected for vehicle $v_i$, then zero or one customer is assigned. Conversely, if no station is assigned, two customers must be assigned. In addition, a vehicle assigned to a station must not utilize the second customer slot. These constraints are formulated as:
\begin{equation}
H_{A_2} = \sum_i\left\{\left(1-\sum_{j}\sigma_{v_i, s_j}\right)\left(2-\sum_{k,t}\sigma_{v_i,c_k,t}\right)+\sum_{j}\sigma_{v_i,s_j}\sum_{k}\sigma_{v_i,c_k,2}\right\}.\label{eq:vrp_ha2}
\end{equation}
Furthermore, to preclude the assignment of multiple stations or multiple customers to the same time slot, we define the penalty Hamiltonian:
\begin{equation}
H_{A_3} = \sum_i\left\{\sum_{j_1>j_2}\sigma_{v_i,s_{j_1}}\sigma_{v_i,s_{j_2}}+\sum_{k_1>k_2}\left(\sigma_{v_i,c_{k_1},1}\sigma_{v_i,c_{k_2},1}+\sigma_{v_i,c_{k_1},2}\sigma_{v_i,c_{k_2},2}\right)\right\}.\label{eq:vrp_ha3}
\end{equation}
The total travel time cost function comprises the costs associated with the three route patterns. Thus, the cost Hamiltonian is:
\begin{equation}
H_{B} = \sum_i\left\{\sum_jt\left(x_{v_i},x_{s_j}\right)\sigma_{v_i,s_j}\left(1-\sum_k\sigma_{v_i,c_k,1}\right)+\sum_{j, k}t\left(x_{v_i},x_{c_k},x_{s_j}\right)\sigma_{v_i,s_j}\sigma_{v_i,c_k,1}+\sum_{k_1,k_2}t\left(x_{v_i},x_{c_{k_1}},x_{c_{k_2}}\right)\sigma_{v_i,c_{k_1},1}\sigma_{v_i,c_{k_2},2}\right\}.
\end{equation}
Finally, the total Hamiltonian is defined as
\begin{equation}
H=H_{A_0}+H_{A_1}+A_2H_{A_2}+H_{A_3}+BH_{B}
\end{equation}
where $A_2$ is a coefficient adjusted to ensure that constraint violations result in a positive energy, and $B$ is the weight for the travel cost (set to 0.001 in this paper). Regarding the coefficient $A_2$, the term $\left(1-\sum_{j}\sigma_{v_i, s_j}\right)\left(2-\sum_{k,t}\sigma_{v_i,c_k,t}\right)$ in Eq. (\ref{eq:vrp_ha2}) can yield a negative value if multiple stations are assigned to vehicle $v_i$. Specifically, the magnitude of this negative energy increases linearly with the number of assigned stations. On the other hand, the violation penalty $\sum_{j_1>j_2}\sigma_{v_i,s_{j_1}}\sigma_{v_i,s_{j_2}}$ in Eq. (\ref{eq:vrp_ha3}) increases quadratically based on the number of combinations. Therefore, the constraint is tightest when the number of assigned stations is minimal, i.e., two. In this case the term in $H_{A_2}$ takes -2, while the penalty in $H_{A_3}$ is 1. To prevent the total energy from becoming negative due to the violation, the condition $-2A_2 + 1 > 0$ must be satisfied. This leads to $A_2 < 1/2$ (we use $A_2=1/3$ in this study).

\section*{Data availability}
The datasets used during the current study are available from the corresponding author upon reasonable request.

\bibliography{qa}

\section*{Author contributions statement}
T.G. conceived of the presented idea and performed the experiments. M.O. verified the analytical methods and supervised the findings of this work. All authors discussed the results and contributed to the final manuscript.

\section*{Competing interests}
T.G. is an employee of Honda R\&D Co., Ltd. and a student at Tohoku University. M.O. is a professor at Tohoku University.
Tohoku University received a research grant from Honda R\&D Co., Ltd. to support this research.
T.G. and M.O. declare no additional potential conflicts of interest.

\section*{Additional information}

\end{document}